\documentclass{elsart}
\usepackage{rotating}
\usepackage{graphicx}
\def\floatsep{1.5cm}		
\def\textfloatsep{0.5cm}	
\def\intextsep{0.5cm}		

\def\Journal#1#2#3#4{{#1} {\bf #2}, #3 (#4)}

\def\NIMA{{\em Nucl. Instrum. Methods} A}

\begin{document}                                                 

\def\bibname{References}
\bibliographystyle{plain}

\begin{frontmatter}

\title{THE TARGET SILICON DETECTOR FOR THE FOCUS SPECTROMETER}
%

\long\def\inst#1{\par\nobreak\kern 4pt\nobreak
    {\it #1}\par\vskip 10pt plus 3pt minus 3pt}
The FOCUS Collaboration\footnote{{See http://www-focus.fnal.gov/authors.html for additional author information.}}
\author[ucd]{J.~M.~Link},
\author[ucd]{M.~Reyes},
\author[ucd]{P.~M.~Yager},
\author[cbpf]{J.~C.~Anjos},
\author[cbpf]{I.~Bediaga},
\author[cbpf]{C.~G\"obel},
\author[cbpf]{J.~Magnin},
\author[cbpf]{A.~Massafferri},
\author[cbpf]{J.~M.~de~Miranda},
\author[cbpf]{I.~M.~Pepe},
% ...
\address[ucd]{University of California, Davis, CA 95616} 
\address[cbpf]{Centro Brasileiro de Pesquisas F\'isicas, Rio de Janeiro, RJ, Brasil} 
\pagebreak
\address[cinv]{CINVESTAV, 07000 M\'exico City, DF, Mexico} 
\address[cu]{University of Colorado, Boulder, CO 80309} 
\address[fnal]{Fermi National Accelerator Laboratory, Batavia, IL 60510} 
\address[fras]{Laboratori Nazionali di Frascati dell'INFN, Frascati, Italy I-00044} 
\address[ui]{University of Illinois, Urbana-Champaign, IL 61801} 
\address[iu]{Indiana University, Bloomington, IN 47405} 
\address[korea]{Korea University, Seoul, Korea 136-701} 
\address[koreak]{Kyungpook National University, Taegu, Korea 702-701} 
\address[milan]{INFN and University of Milano, Milano, Italy} 
\address[nc]{University of North Carolina, Asheville, NC 28804} 
\address[pavia]{Dipartimento di Fisica Nucleare e Teorica and INFN, Pavia, Italy} 
\address[pr]{University of Puerto Rico, Mayaguez, PR 00681} 
\address[sc]{University of South Carolina, Columbia, SC 29208} 
\address[ut]{University of Tennessee, Knoxville, TN 37996} 
\address[vu]{Vanderbilt University, Nashville, TN 37235} 
\address[wisc]{University of Wisconsin, Madison, WI 53706} 

%

\begin{abstract}

We describe a silicon microstrip detector interleaved with segments of
a beryllium oxide target 
which was used in the
FOCUS photoproduction experiment at Fermilab. The detector was designed 
to improve the vertex resolution and to enhance the reconstruction
efficiency of short-lived charm particles. 
\end{abstract}

\end{frontmatter}

\section{Introduction}

%
%
%

FOCUS is a Fermilab fixed target photoproduction (Average $E_\gamma\sim180$ GeV) experiment which has
been configured to investigate the production and decay of charmed particles.
The FOCUS experiment is an upgraded 
version of Fermilab
experiment E687~\cite{e687_nim,e687_beam}. One of the significant
improvements made to the
E687 detector is the segmentation of the target and the installation of 
four supplementary silicon planes called the Target Silicon Strip Detector (TSSD).

In photoproduction, relatively long lived charm particles are principally identified by 
the separation between the primary production point within the target material 
and the secondary 
decay vertices of the charm particle and the anti-charm particle. Generally, only one of the
two secondary vertices is identified.
Secondary vertices which are identified outside of the target material are mainly due to  
particle decays. Inside the target material secondary 
vertices can result from secondary hadronic interactions as well as from decays. 
By searching for decays outside of the
target material there is a significant improvement in signal to noise in charm particle
mass plots. We used a target composed of beryllium oxide (BeO)  because of its
relatively  high
density, ~3.0 gm/cm$^3$, which allows for a compact target, and its low number of 
atomic electrons, which reduces the number of radiation lengths per interaction
length. The principal drawback to a segmented target arrangement is that the target interaction
region grows in length. If the only tracking measurements are  made downstream of the 
target, then the spatial resolution of vertices is degraded as the
decays occur more and more upstream.  This loss in resolution can be reduced
if additional silicon strip measurements are made inside the target region.

\begin{figure*}
\begin{center}
{\includegraphics[width=7cm,angle=270]{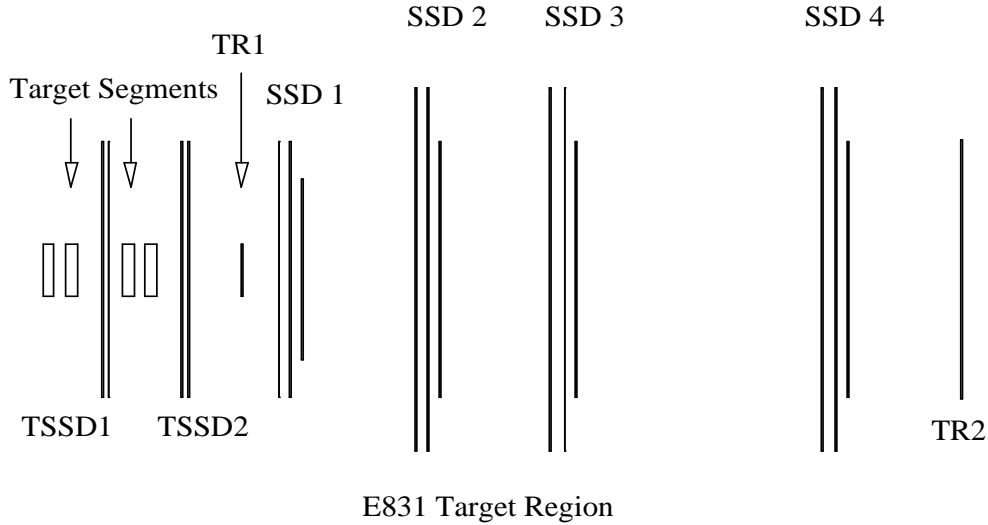}}
\end{center}
\caption{
View of the Target region configuration of the FOCUS spectrometer.
The SSD's are the E687 silicon microvertex detector planes,
TSSD's are the Target silicon, there are 2 scintillator triggers (TR1 and TR2),
and 4 Beryllium Oxide targets.} 
\label{fig:spec97}
\end{figure*}

\begin{figure*}
\begin{center}
{\includegraphics[width=5.5in]{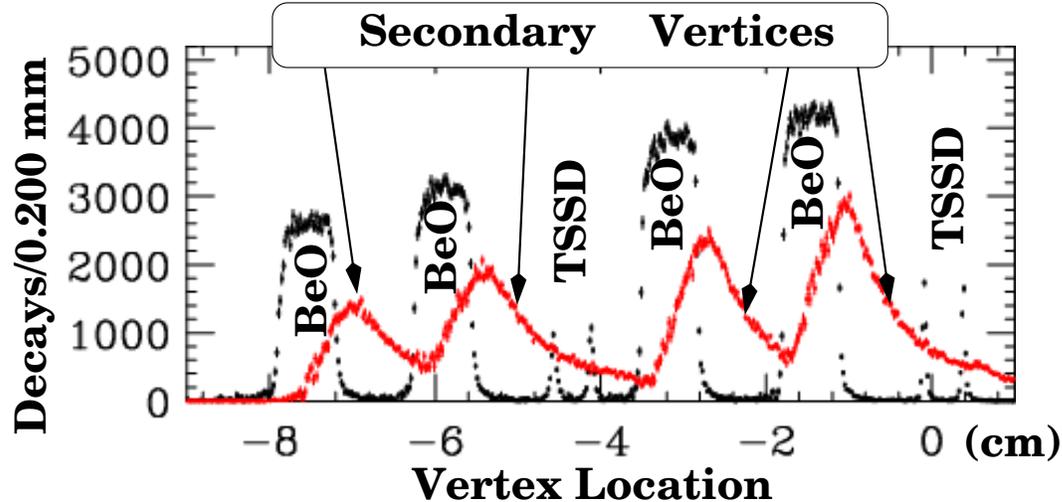}}
\end{center}
\caption{
The location of primary and secondary vertices from charm mesons reconstructed
using the FOCUS reconstruction code. In this figure, the primary vertex locations
mimic the locations of the target material and TSSD planes as shown in
FIG. \ref{fig:spec97}. Note also the large number of secondary vertices that
occur outside of the target and/or TSSD material.} 
\label{fig:outside}
\end{figure*}

The FOCUS target region layout is schematically displayed in Fig. \ref{fig:spec97}
and reconstructed vertices from charm are shown in Fig. \ref{fig:outside}.  
Each of the four BeO target segments  is 3.0~cm square and 6.75~mm long. A 1.0~cm 
decay region follows each target segment.  Two sets of TSSD plane doublets
are located behind the second BeO target and behind the most downstream BeO target.
The transverse size of the BeO target segments was chosen to match the photon 
beam profile.  The TSSD planes are followed by a scintillator (TR1) used for triggering,
four stations of Silicon Strip Detectors (SSD's) with each station consisting
of three silicon microvertex planes, and a second scintillator, (TR2). By demanding  a hit
in the TR1 scintillator we attempt to select interactions that occurred 
in the target region. A coincidence hit in TR2 reduces the number of spurious 
noise hits in TR1.  The main vertex tracking is accomplished in the 
SSD's and is described in detail in reference 3.

The components of the Target Silicon Strip Detector
consist of 4, 25 micron pitch, 300 micron
thick planes of silicon microstrip detector and the electronics necessary to
amplify, digitize, and store the information.
In Section 2 we describe the physical layout of a Target
Silicon Strip plane and how the planes and target segments were
mounted.
In Section 3 we describe the readout electronics.
In Section 4 we 
discuss how the hit information in the TSSD's is matched up with the 
tracking information from the downstream SSD planes.
In Section 5 we describe the performance of the TSSD. In section 6 
we detail the use of the detector as a target rather than as an 
active detector.

\section{Physical Description of the TSSD's}
 
\begin{figure*}
\begin{center}
{\hskip 1.0cm{\includegraphics[width=10cm,angle=90]{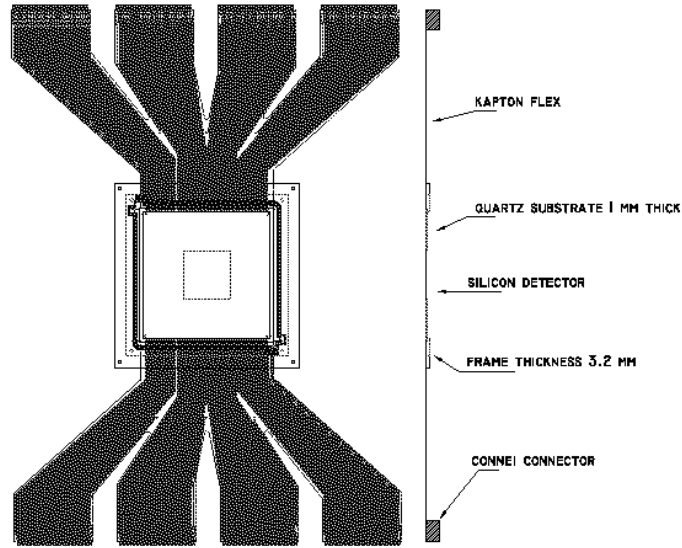}}}
\end{center}
\caption{
Front view and side view of a single target silicon strip detector 
plane and fanout. The strips travel vertically in the figure.}
\label{fig:ribbonk}
\end{figure*}

Each target silicon wafer of 2048 channels has an active area of 
5~cm square. The silicon wafer is 
wire-bonded onto a 15~cm square, 1~mm thick quartz plate. The central 
1024 channels of
the silicon wafer are fanned out on the quartz plate and wire-bonded to 
two flexible ribbons made of very thin circuit board material. 
The outer 1024 (512 on a side)  channels are tied 
together and grounded making
the effective active area of each plane  2.5~cm $\times$ 5~cm.
The outer edge of each flexible ribbon is attached to a 132 pin Connei connector.
A front and side view of a single plane is shown in Fig. \ref{fig:ribbonk}.
The planes were fabricated by MICRON SEMICONDUCTOR Limited~\cite{wilburn} in 
England and were
adapted from a WA85 Chip design. In order to provide greater rigidity and to
allow for alignment the quartz plate is mounted onto a G-10
frame.  Finally, two G-10 frames from a doublet are mounted onto a holding fixture. 
The holding fixture is held by bushings on low carbon steel rods which 
are fixed to a granite mounting. The segmented targets and the
two doublets are supported by the steel rods. Precise location along the
beam direction is performed using spacers on the steel rods. Finally, the planes
and support structures were mounted inside a Faraday shielded, temperature controlled hut. 

The two planes in each doublet are aligned at $\pm$45 degrees to the horizontal.
Each plane has a coordinate in the same view as one of the views in the
external SSD silicon array. The ribbons from each doublet of detectors are
attached to an aluminum housing which has a separate connection to the
granite support. The aluminum housing holds the individual Connei connectors and   
supports the front-end electronics. The advantage of this arrangement is that
the front-end electronics support is disconnected from the individual silicon
plane support and allows for the removal of the pre-amplifiers without
damaging the wafer alignment.  

We purchased five planes from MICRON SEMICONDUCTOR Limited in order to have 
one spare plane. All planes had a 
typical strip breakdown voltage in excess of 100 V. Other electrical 
properties for the five planes are presented in Table \ref{tab:ssdtab}.
We operated the planes with higher bias voltages (over-depletion) than 
the factory specified full depletion voltage listed in Table \ref{tab:ssdtab}.

\begin{table*}
\caption{Characteristics of the Target Silicon Strip Detectors.
Columns four, five, and six are typical values for the resistance
between strips, strip leakage 
current at the depletion voltage and at the depletion voltage + 20 volts.
\label{tab:ssdtab}}
\vspace{0.4cm}
\begin{center}
\begin{tabular}{ccccccl}
\hline
 Plane \#  & Thickness & Depletion & Resistance & Leakage & Leakage Current\\
           &  ($\mu$m) &   Voltage & Between Strips & Current & at Dep. + 20 V\\
\hline
    1      &  303      &  10.0 & 3 M$\Omega$ & 10nA        & 20 nA   \\
    2      &  301      &  10.8     & 5 M$\Omega$ & 9nA         & 17 nA   \\
    3      &  287      &  19.5     & 13 M$\Omega$ & 4nA        & 6 nA    \\
    4      &  303      &  25.0       & 20 M$\Omega$ & 9nA        & 13 nA   \\
   (spare) &  291      &  20.1     & 40 M$\Omega$ & 2nA        & 5 nA    \\
\hline
\end{tabular}
\end{center}
\end{table*}

\section{Readout Electronics of the TSSD's}

\begin{figure*}[htpb]
\begin{center}
{\includegraphics[width=14cm]{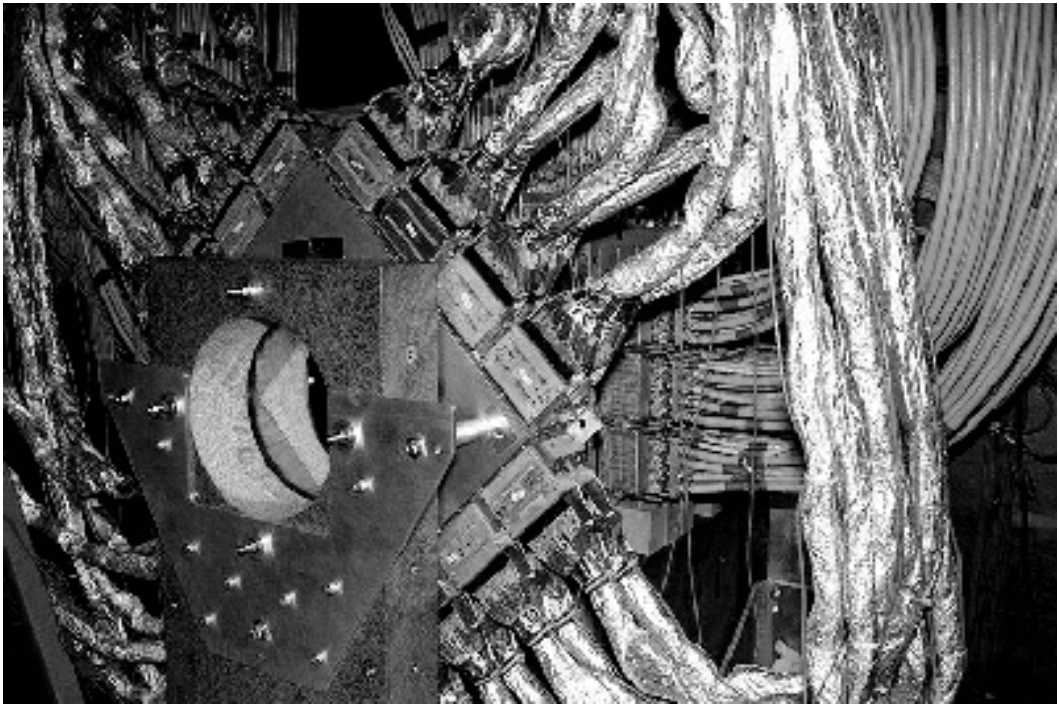}}
\end{center}
\caption{The Target silicon system as installed in experiment E831. 
Note the granite support structure and the boxes containing MSP1 amplifiers 
that are connected to the shielded ribbon cable bundles. In the
background, the SSD pre-amplifier boxes and cables are clearly visible.}
\label{fig:conimg}
\end{figure*}

The signals from each of the individual silicon strips are fed through Connei
connectors into MSP1 pre-amplifiers~\cite{Avondo}.  The amplifiers are 
mounted on the aluminum 
support housing as shown in Fig. \ref{fig:conimg}. 
The  MSP1 pre-amplifiers have an intrinsic 20~ns rise time 
and about a 100~ns wide response to a 7~ns wide square pulse and
have a good noise figure even in the presence of a high input
capacitance (50~pF).\footnote{Both the added capacitance of a detector 
and resistance in the readout serve to increase the noise
coming from a pre-amplifier.} 

The differential output
signals from each group of MSP1's are sent over 40~ft. long twisted pair 
cables that are bundled together and shielded with aluminum foil and 
clear packing tape. These bundles are then split into groups of 64 channels
and read into electronics boards (AMP/DIG) that amplify and digitize the analog
signals coming from the MSP1's.

\begin{sidewaysfigure*}[htpb]
\begin{center}
{\includegraphics[width=22cm]{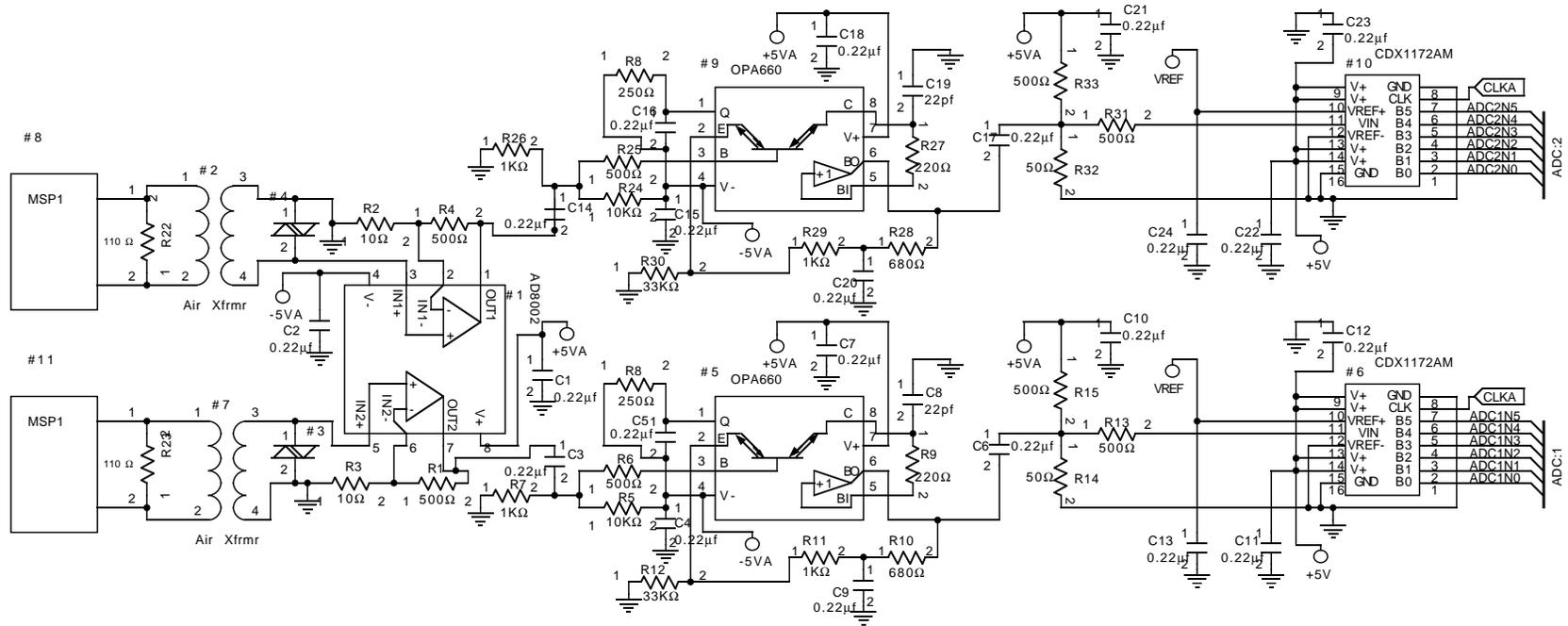}}
\end{center}
\caption{
AMP/DIG board schematic for 2 channels from the MSP1 inputs to the 6 bit
flash ADC output.}
\label{fig:chnl}
\end{sidewaysfigure*}

In the AMP/DIG boards, the signals pass through a 1:1 transformer 
to enhance isolation from the upstream electronics, and then are amplified 
by a factor of 50 using an AD8002 amplifier.  The charge is integrated
using an OPA660 diamond transistor as a non-feedback ns-integrator~\cite{burrb}
to produce a voltage level for the CDX1172AM FADC chip (see Fig. \ref{fig:chnl}). 
This voltage is sampled every 54~ns 
(3 accelerator buckets) and stored in a buffer memory. 
The RC decay time for the signal was chosen 
to be about 800~ns to lessen the variation of signal size for the 3 
accelerator buckets covered by each sample.

%
%
%
 
When the experimental trigger fires, two 
signals in the buffer memory, one {\it{in--time}} and one 3 samples
($\sim 162~{\rm{ns}}$) early, are fed
into an arithmetic logic unit (ALU). The earlier sample
is subtracted from the {\it{in--time}} sample.\footnote{The baseline subtraction helps
reject low frequency common-mode noise as well as reduce the
effects of pile-up.} If the difference exceeds the threshold
stored in the ALU for that channel, the FADC information is 
given a channel address and stored in an output
buffer. Output buffers are read out sequentially across a crate
of AMP/DIG boards.

The AMP/DIG boards are organized so that each silicon plane can
be read out by one crate of electronics. Each crate holds 16 9U$\times$280~mm
VME boards. Constants for each board are downloaded from a CAMAC
controller board specially designed to both control a crate of
AMP/DIG boards and organize the information from a crate into
an event record. Each event record consists of 16 bit data words,
a crate identification word, and a word count. The event records 
from all 4 crates are transmitted from the experimental hall
to the counting house at ECL levels on a single twisted flat cable. 
Typically, all 4 crates are read out in less than $15~\mu {\rm{s}}$. 
An average event record for a target silicon plane due to a 
charm containing event consists of
about 6 channels of electronics noise which exceeds threshold,
an additional 7 channels due to $e^+e^-$ pair contamination, and 
14 channels due to the particles created from the charm. 

\begin{figure*}
\begin{center}
{\includegraphics[width=14cm]{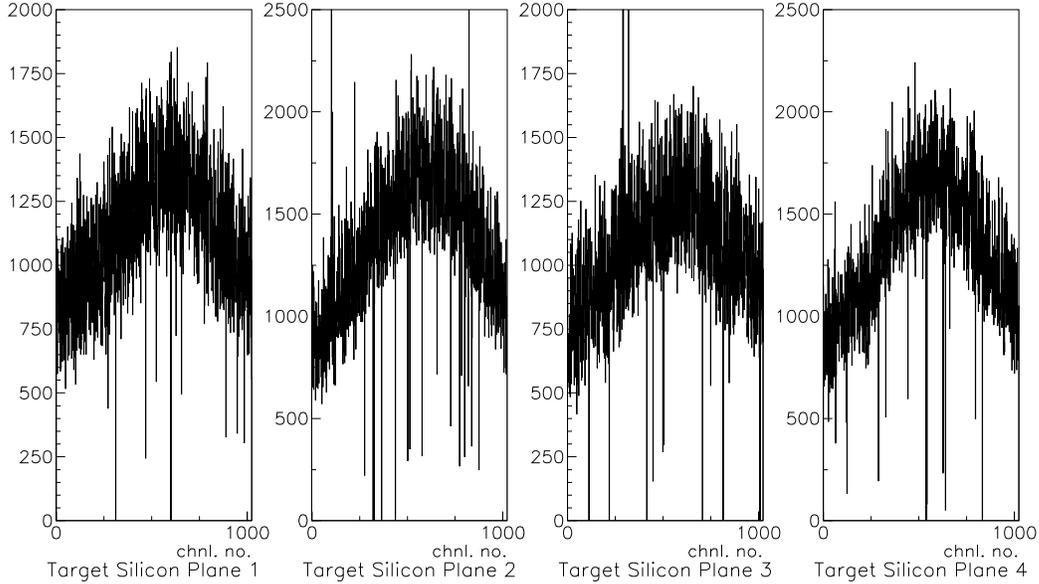}}
\end{center}
\caption{
Profile of hits in for each target silicon plane for a sample of photon data.
Planes are ordered 3, 1, 4, 2 from most upstream to most downstream.}
\label{fig:profil}
\end{figure*}

The profile of hits from regular photon data is presented in 
Fig. \ref{fig:profil}. The number of missing and inefficient channels
is of order 1-2\%. Due to charge sharing, a missing channel does not 
necessarily translate to a loss in efficiency but rather a loss in
resolution.

\section{Hit Reconstruction}

To incorporate the TSSD's in the track reconstruction, we refit
existing
SSD tracks in a staged process. The information from 
the closest (most downstream) planes of the TSSD is used first. Once
refit with any available information from the downstream target silicon
planes, the
tracks are then extrapolated to the most upstream planes and
refit again. All three categories of tracks, the original SSD track,
the tracks refit using the downstream TSSD information (if any), and the
tracks refit using information from all TSSD planes (if any) are
saved for later vertex determination. This approach allows us to include
multiple coulomb scattering effects in a natural way, use the well tested
SSD tracking algorithm in a seamless way, and find the best solution for
the refit track at a point nearest the decay and/or production vertex.

To begin the process, the SSD tracks are first refit taking into
account multiple coulomb scattering in such a manner as to find
the best fit for tracks {\it{entering}} the SSD system.\footnote{
SSD Tracks
are originally fit to optimize momentum determination, i.e. to 
find the best parameters for tracks {\it{leaving}} the SSD system.}

The
refit SSD tracks are extrapolated to the two most downstream target
silicon planes and a three sigma search radius (about 1-2 strips typically) 
is identified to
search for hits. The search is not allowed to extend for more
than $\pm~40$ strips.  

\begin{figure*}
\begin{center}
{\bf {\includegraphics[width=14cm]{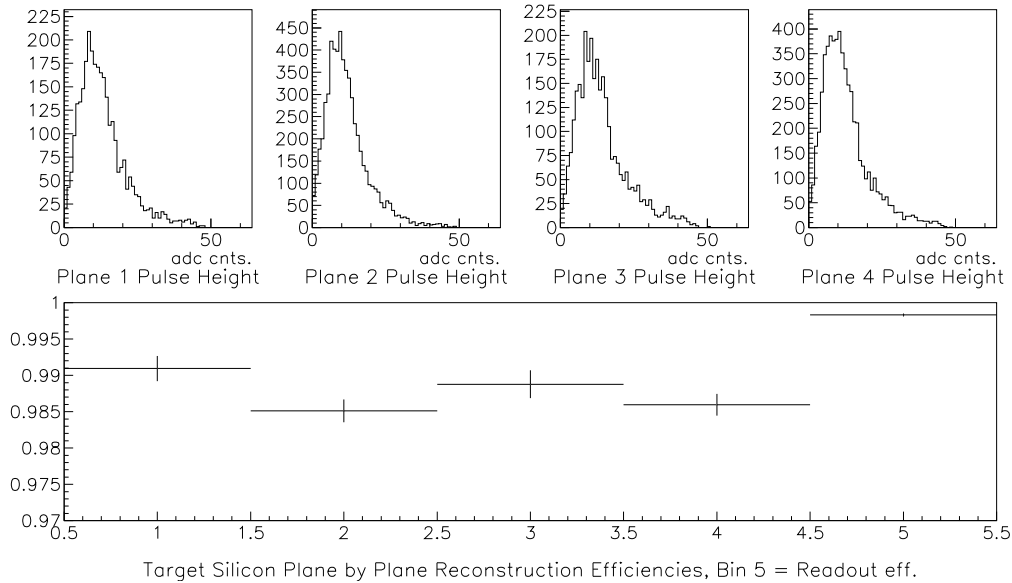}}}
\end{center}
\caption{
The ADC pulse heights (pedestal is subtracted in the electronics) 
for hits used to refit a track.
The  individual plane efficiency
is calculated for high momentum tracks on a run by run basis.
We show these efficiencies in the lower plot. The fifth bin
gives an estimate of how often (about $0.1\%$ of the time) 
we must discard an event due to
high multiplicity in a single plane or readout errors since it
records the efficiency for retaining an event.}
\label{fig:ADC}
\end{figure*}

The closest non-zero (zero count ADC hits are hits that just exceed the 
baseline subtraction) ADC hit to an extrapolation is used as a seed to
determine the number of adjacent strips that fired for the extrapolated
SSD track.  Once the hits are determined, a simple linear weighting
between the seed hit and the adjacent hit with the largest signal is used 
to determine the
centroid. If more than 2 hits are found, none of the hits are used in
the refit, although the number of hits, the summed ADC information, and
the centroid are saved. The ADC pulse height distribution for the hits
found during the refit to the extrapolated SSD tracks is plotted in Fig.
\ref{fig:ADC}. A minimum ionizing peak is clearly observed in all planes.

To further reduce confusion
the extrapolations of other SSD tracks are checked to see if there 
is overlap with the SSD track used to find the seed hit. No hits
are allowed to be shared in the most upstream planes, but 
single hits in each downstream plane are allowed to be shared with
another SSD track. The number of ADC hits used, and the number 
of overlapping SSD track extrapolations are used to determine
the resolution error used in the refit. 

At the end of this refit process, we use the SSD tracks and the
E687 free form  vertexing method~\cite{e687_nim} to
determine the most likely location of the production vertex.
If this vertex occurs upstream of one or both of the TSSD
doublets, SSD tracks are replaced with tracks refit using the
TSSD. Finally, the vertex location is used to calculate the 
amount of multiple coulomb scattering to include in the analysis
of the event. Further refinements are possible, such as for neutral 
kaons that decay relatively far from the production vertex, but 
seldom necessary for charm reconstruction. 


\section{TSSD Performance}

The FOCUS experiment took data during the 1996-1997 fixed target
run at Fermilab from September 1996 to September 1997. 
The TSSD was installed in the middle of data taking
during the holiday (Dec.--Jan.) 
shutdown of the Fermilab Tevatron. The information from the  
TSSD is present for about $70\%$ of the FOCUS data.
For a short period (less than $10\%$ of the data
containing TSSD information) after the installation,
plane number 1 had readout problems that resulted in a loss of 
resolution and some efficiency for this plane. Otherwise, the
TSSD performed consistently at high efficiency for
the duration of its use. The reconstructed data, as well as
the simulation of the TSSD, have been thoroughly
tested and utilized in all published FOCUS physics analyses 
to date
\cite{2002ge,2001ee,2001qy,2001zj,2001hz,2001rn,2000kr,2000aw}. 
With the additional information from 
the TSSD we significantly improved tracking resolution in the 
target region of the FOCUS spectrometer. The improvement is 
demonstrated in several tests using a representative 
subsample of the data.

\begin{figure*}
\begin{center}
{\includegraphics[width=14cm]{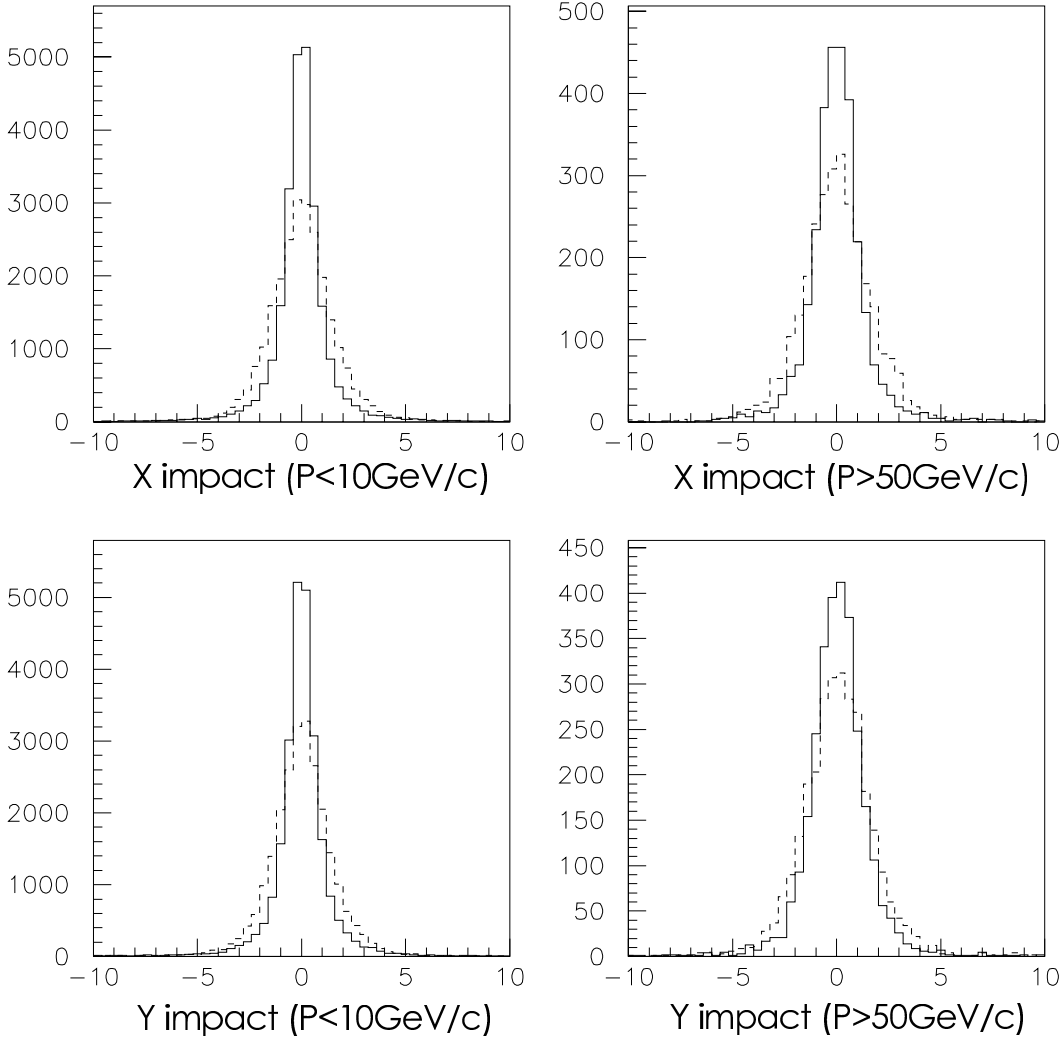}}
\end{center}
\caption{
X and Y (transverse)
normalized impact parameters. 
The dashed line represents the standard
vertex reconstruction algorithm, the solid line is after including
target silicon information into the fit(the error estimate is
unchanged on purpose so that the improvement in resolution is evident).
Both low momentum ($P<10~GeV/c$) and high momentum ($P>50~GeV/c$)
tracks are shown for both transverse (X and Y) projections.}
\label{fig:RESOL}
\end{figure*}

In Fig. \ref{fig:RESOL} we show the normalized transverse 
impact parameters in two views and for two momentum regions. 
Vertices with three or more tracks are refit after removing one of the
tracks. This removed track is
projected back to the reconstructed vertex and the transverse miss 
distance between the track and the vertex determined and normalized
using the computed error for the miss. In order to
show the improvement in resolution using the TSSD, the same error 
estimate is used for the case with and without the TSSD included
in the tracking information. It is clear that using the TSSD information
improves the position resolution.

\begin{figure*}
\begin{center}
{\includegraphics[width=8cm]{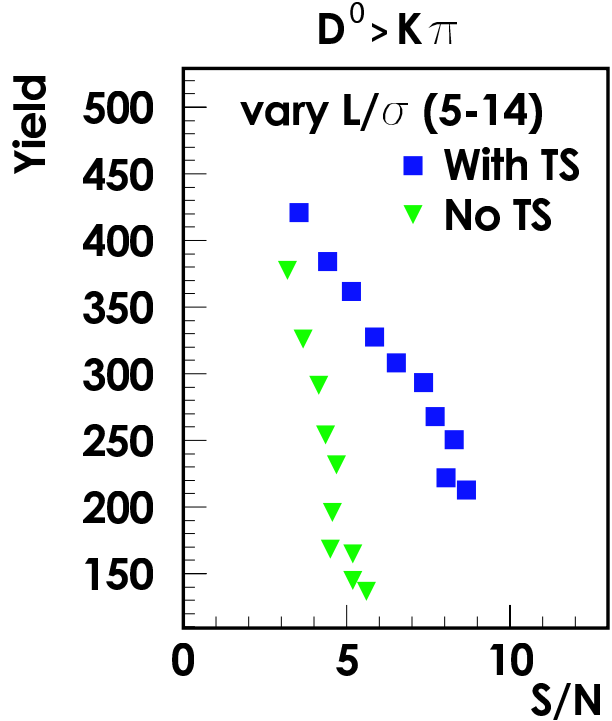}}
\end{center}
\caption{
Yield versus the Signal to
Noise Ratio (S/N) from an
 analysis using the same data of the decay mode
$D^0\rightarrow K^{-}\pi^+$. The analysis was
done utilizing SSD alone (``no TS'') and a combination
of the SSD  and Target Silicon systems (``With TS'').
Each point represents a different value of the statistical significance
of the detachment $(L/\sigma_{L})$
of the $K^{-}\pi^+$ vertex from the primary interaction vertex.
Notice the improvement in (S/N) as we increase $(L/\sigma_{L})$ from 5 to 14. 
At the largest $(L/\sigma_{L})$, we have $\sim 50\%$ more yield and $\sim 40\%$
less background when we use the Target Silicon Strip Detector in the analysis.}
\label{fig:matt}
\end{figure*}

In Fig. \ref{fig:matt} we present a  plot of the charm yield for the
decay $D^0\rightarrow K^-\pi^+$ using the SSD system alone and then in
conjunction with
the TSSD system. By making an exponential 
fit to drop off in yield as a function of detachment we find that we have 
improved lifetime resolution by $35\%$. It is obvious that the charm decays 
have a better signal/noise ratio with higher yields for the same significance of
separation. 

\begin{figure*}
\begin{center}
{\includegraphics[width=14cm]{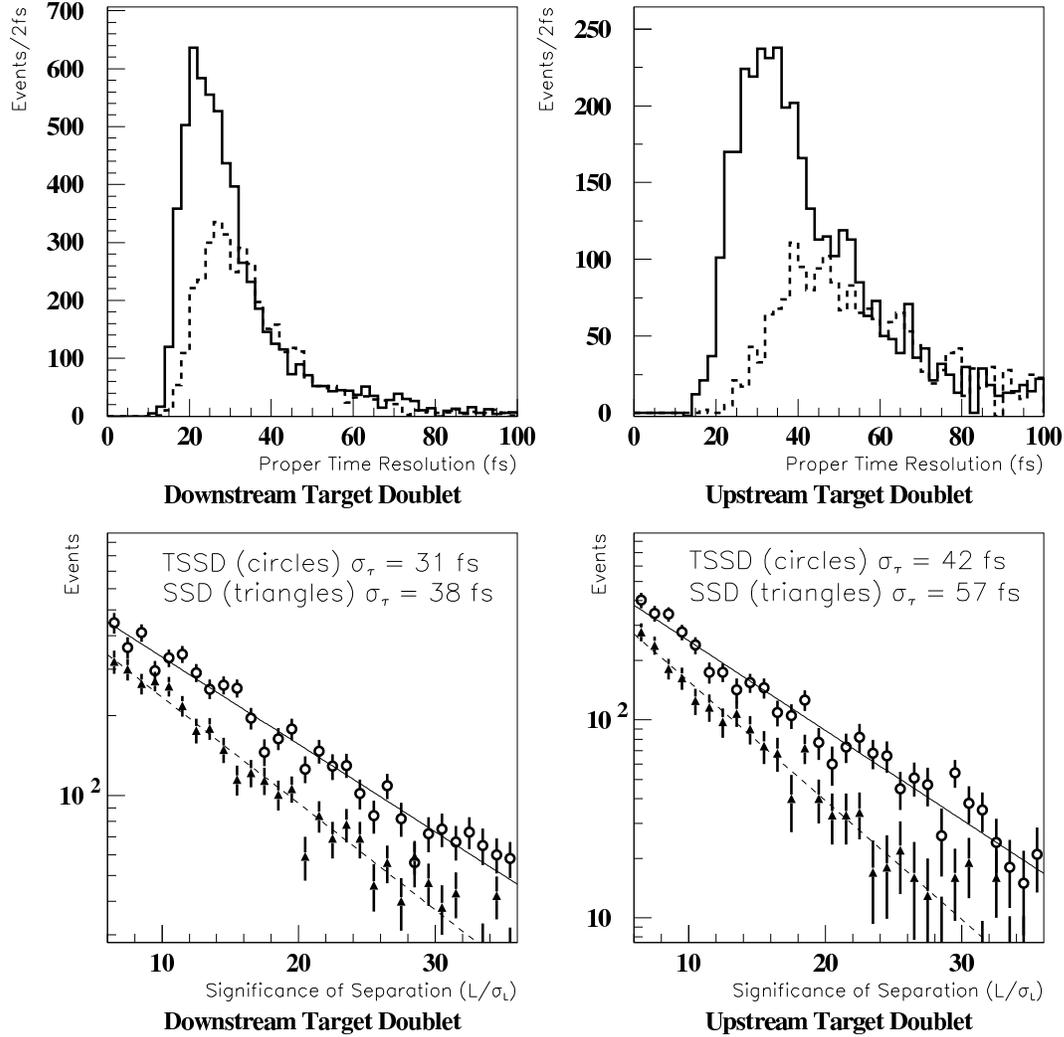}}
\end{center}
\caption{
The calculated proper time resolution from an analysis of the decay mode
$D^0\rightarrow K^{-}\pi^+$. The analysis was
done with data from a period of the FOCUS running when the
TSSD was installed and used in the reconstruction (solid
lines) and when only the 4 BeO targets were present (dashed
lines). The data is split into events from the most downstream
(closest to the SSD) and the most
upstream (farthest from the SSD) pair of Be0 targets.
The distributions are identically normalized by eye at
large resolutions. From the top two plots, one sees that using the 
target silicon information in the reconstruction effectively 
increases the fraction of events with very good proper time resolution.
In the bottom two plots, we performed a fit to the evolution of 
$(L/\sigma_{L})$ to quantify the resolution improvement. The TSSD
information improved the average proper time resolution by about $20\%$.
}
\label{fig:jimres2}
\end{figure*}

In Fig. \ref{fig:jimres2} we present a  background subtracted plot of the proper 
time resolution and the $(L/\sigma_{L})$ evolution for the 
decay $D^0\rightarrow K^-\pi^+$ using a 
representative subset of the data from 
different running periods of the FOCUS experiment. During the running
without the TSSD planes installed, the Be0 targets were moved about
1 cm closer to the SSD system. In addition to compensating for the shift
of the targets away from the SSD, using the target silicon information 
significantly enhances the proper time resolution.

\begin{figure*}
\begin{center}
{\includegraphics[width=14cm]{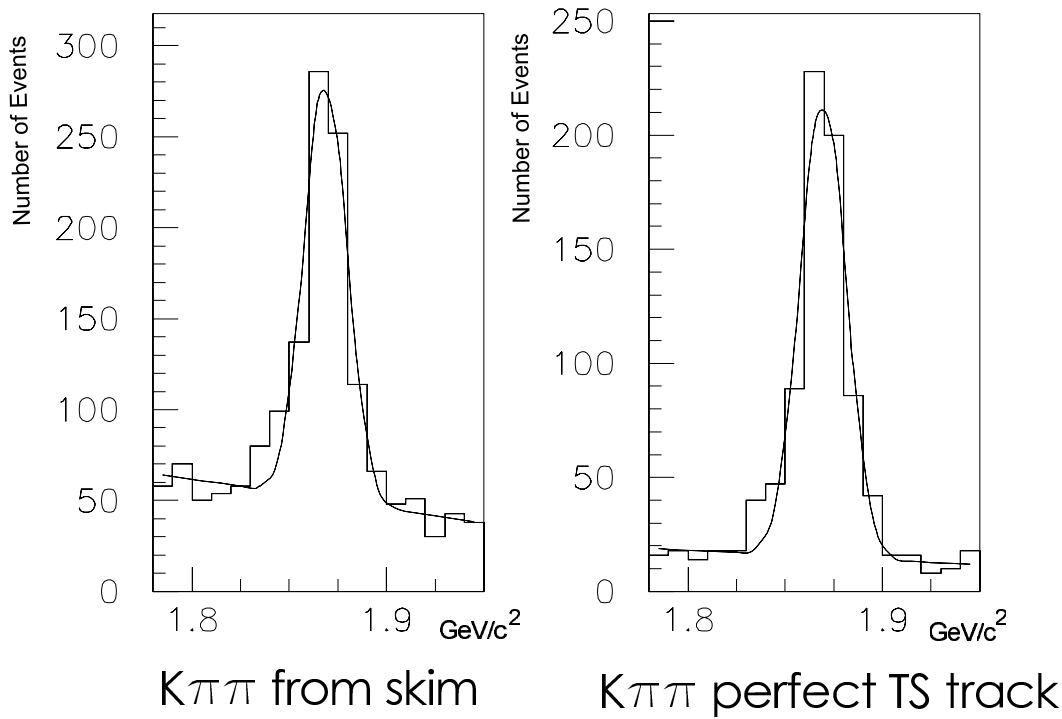}}
\end{center}
\caption{
Histograms of the mass
distribution from a non-optimized analysis of the
decay mode $D^+\rightarrow K^{-}\pi^+\pi^+$. Decay candidates
were selected (``skim'') if the reconstruction indicated that the
$D^+~track$ passed through one of the Target Silicon doublets.
The $D^+~track$ hypothesis was then tested using Target Silicon
information, and if the hypothesis was confirmed, the
candidate was selected for the second plot (``perfect TS track'').
We retain $93\%$ of the signal and remove $70\%$ of the background
by requiring confirmation of the $D^+~track$ hypothesis using
the Target Silicon information.} 
\label{fig:TSILICON4}
\end{figure*}

We have developed  
software to reconstruct long lived $D^+$ mesons {\it{tracks}}
using the charm production vertex and
information from the target silicon planes. In Fig. \ref{fig:TSILICON4}
we present two mass distributions of $D^+\rightarrow K^{-}\pi^+\pi^+$
where the primary and secondary vertex information indicated that the
$D^+~track$ passed through one of the Target Silicon doublets. It is
evident that the signal to noise is substantially improved when 
confirming TSSD hits are found. 

\begin{figure*}
\begin{center}
{\includegraphics[width=14cm]{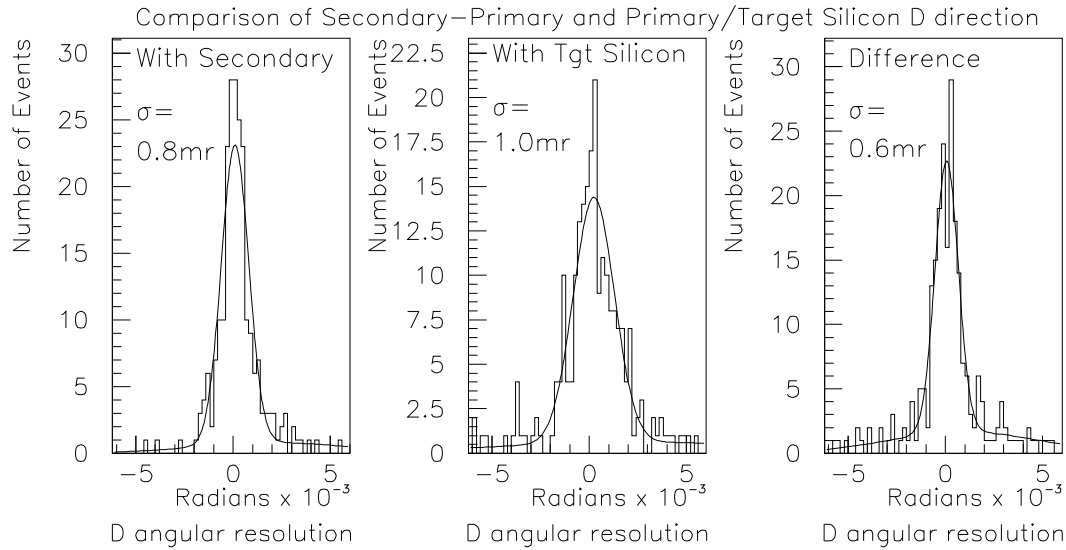}}
\end{center}
\caption{Angular resolution using the vector connecting the primary
and secondary vertex locations of a $D^+\rightarrow K^{-}\pi^+\pi^+$ decay
(left) compared to the resolution using the single prong technique
described in the text (center). In the difference between the two 
techniques (right) we see that the resolutions are highly correlated.
} 
\label{fig:dmu}
\end{figure*}

Tracks from $D^+$ candidates have also 
been reconstructed using just the $D^+$ information in the silicon
and a single track from the decay vertex. Such a technique could be
used for
determining the $D^+$ direction for a form factor analysis of the
decay $D^+\rightarrow K_s^o\mu^+\nu_\mu$. In this instance, the
$D^+\rightarrow K^{-}\pi^+\pi^+$ decays are used again to test the
technique. Briefly, putative D tracks are formed using the primary
vertex and target silicon hits. The D track with the highest 
confidence level to form a vertex with the single track from the
decay vertex is retained and refit using adjacent hits (if any)
in the target silicon. In Fig. \ref{fig:dmu} we show that the resolution 
of the D direction with this technique is comparable to a more 
traditional approach using the vector connecting a fully reconstructed decay 
vertex to the primary vertex. Since the resolutions of the two
methods are highly correlated, it should be possible to make improvements,
such as a better determination of the primary vertex (see Section 6 below),
that benefit both techniques. 

\section{Target Silicon Planes as Target Elements}

\begin{figure*}
\begin{center}
{\includegraphics[width=6.6cm]{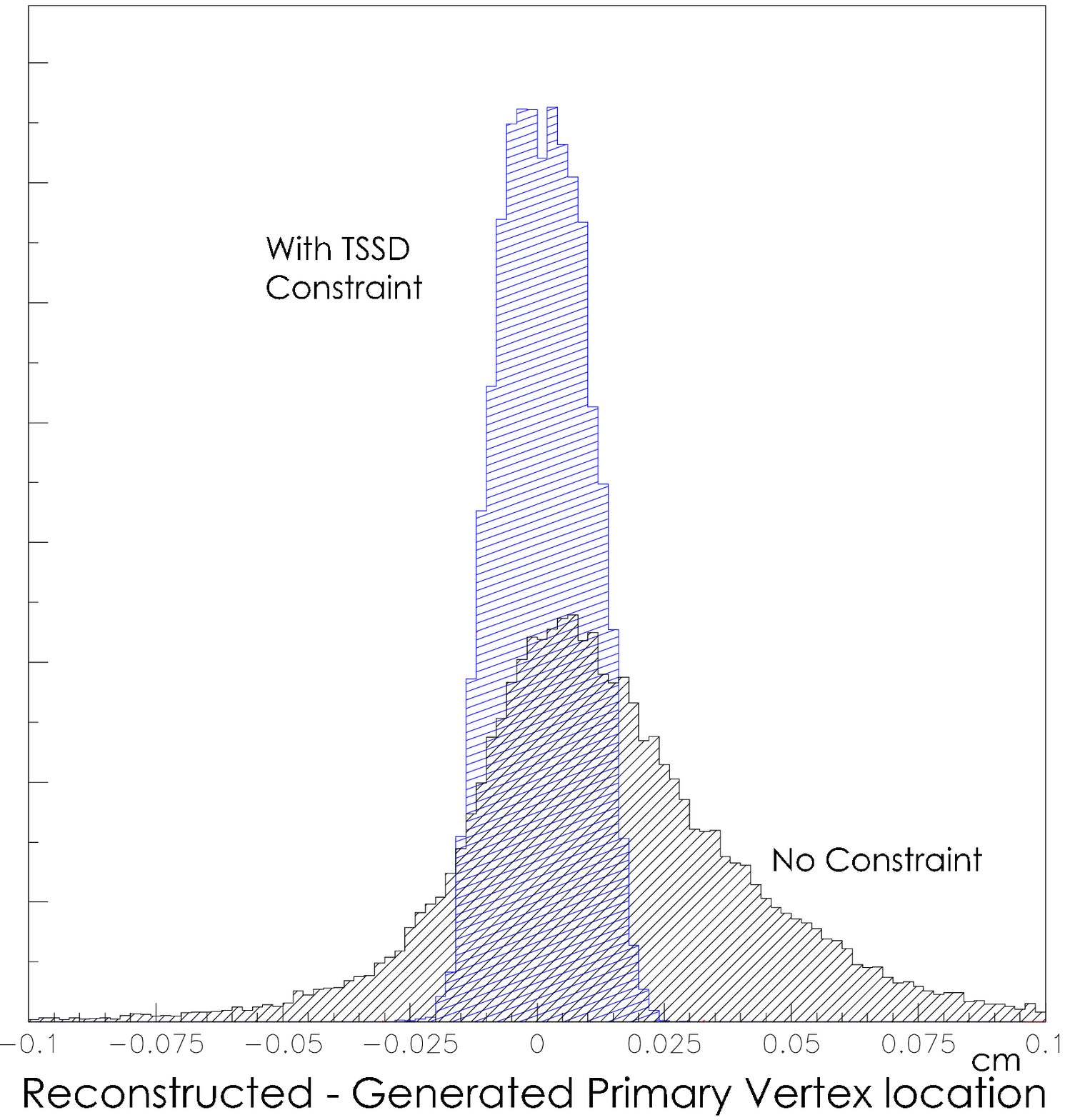}~~\includegraphics[width=6.6cm]{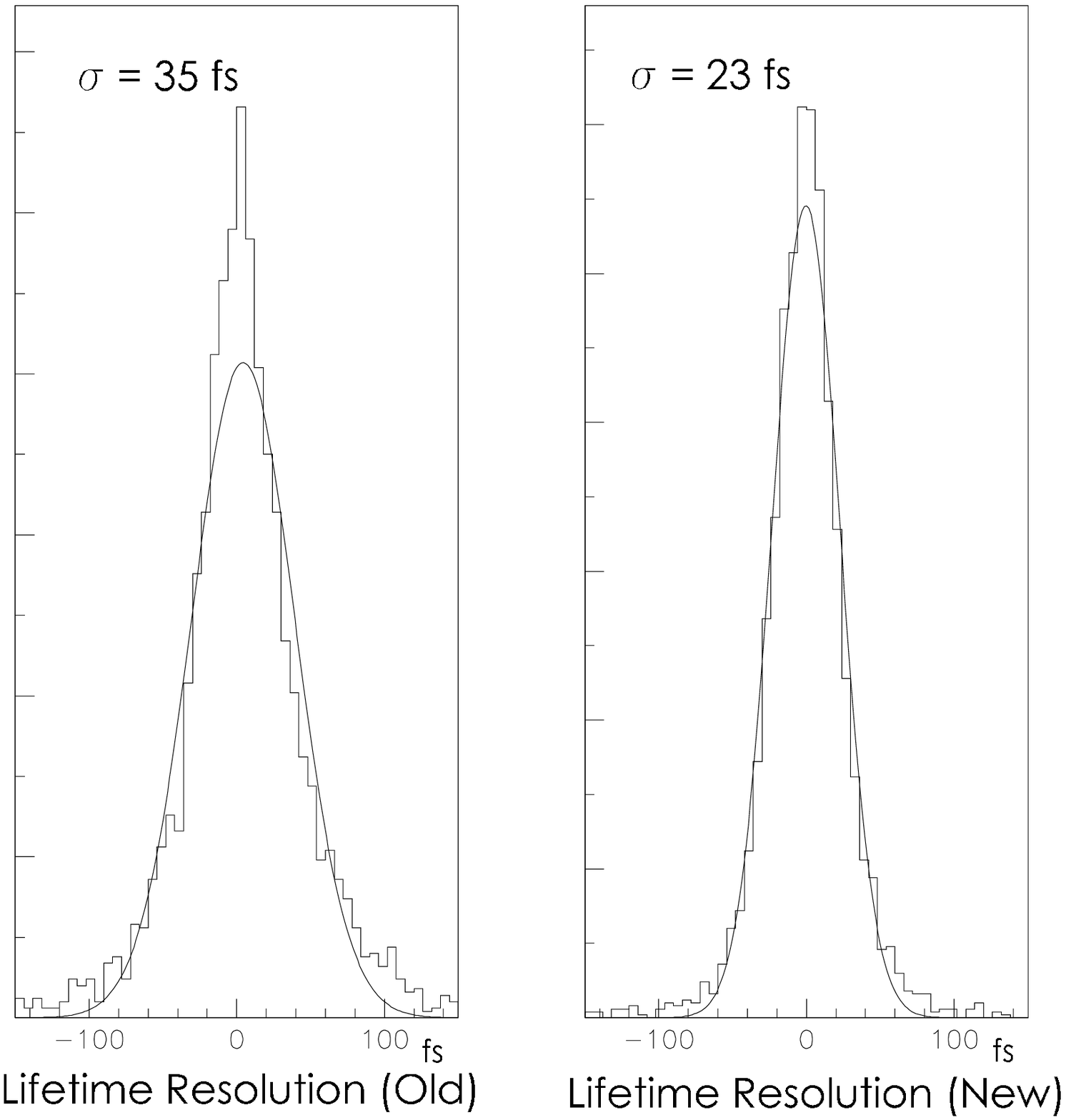}}
\end{center}
\caption{Including the target silicon as a constraint in primary vertex reconstruction
dramatically increased the precision along the beam direction of our estimate 
(left) in the simulation. Proper time resolution for simulated
$D^o\rightarrow K^-\pi^-\pi^+\pi^+$ decays increased from 35 fs (center) to
23 fs (right) using the downstream target silicon doublet only in the constraint.
} 
\label{fig:helms}
\end{figure*}

One way to improve both the direction resolution of a partially 
reconstructed charm decay, and the lifetime resolution for a fully reconstructed 
charm decay, is to improve the resolution of the primary vertex.
Since two hadrons containing
charm are created at the primary vertex,
tracks associated with a partially reconstructed decay are sometimes 
included in the reconstruction of the primary vertex. Due to
the finite lifetime of the charm hadrons involved,
the primary vertex tends to be reconstructed downstream of its
true location. One way to reduce this bias is to use a very
thin target and constrain the primary vertex to be included
in the target material during reconstruction. Using a full simulation
of the E831 spectrometer, we have studied
the improvement possible if one uses the target silicon planes
as passive {\it{targets}} rather than a detector. In Fig. \ref{fig:helms} we show the
results of the technique.
In our simulation, we find that the pull in the primary vertex location 
is reduced and the lifetime resolution substantially improved when
we employ this technique. 

\section {\bf Summary and Conclusions}
    
We have briefly described one of the first silicon strip
systems to be used inside of an experimental target. The use
of the system significantly improved our lifetime and
vertex resolution.  We believe that our experience will
prove useful to future experiments. One can imagine, for instance, 
a future fixed target charm experiment employing many silicon 
(or active diamond) planes in the target region to
take advantage of the techniques presented in this paper.

\section*{Acknowledgments}

We wish to acknowledge the assistance of the staffs of Fermi
National Accelerator Laboratory, the INFN of Italy, and the physics
departments of the collaborating institutions. This research was 
supported in part by the U.~S.
National Science Foundation, the U.~S. Department of Energy, the
Italian Istituto Nazionale di Fisica Nucleare and Ministero
dell'Universit\`a e della Ricerca Scientifica e Tecnologica, 
the Brazilian Conselho Nacional de
Desenvolvimento Cient\'{\i}fico e Tecnol\'ogico, CONACyT-M\'exico,
the Korean Ministry of Education, and the Korean Science and 
Engineering Foundation.

\end{document}